\documentclass[iop]{emulateapj}
\usepackage{graphicx}
\usepackage{apjfonts}
\slugcomment{Accepted for publication in The Astrophysical Journal Letters} 

\begin{document}

\title{OBSCURATION BY GAS AND DUST IN LUMINOUS QUASARS}

\author{S.M.~Usman\altaffilmark{1}}
\author{S.S.~Murray\altaffilmark{1,2}}
\author{R.C. Hickox\altaffilmark{3}}
\author{M. Brodwin\altaffilmark{4}}

\altaffiltext{1}{Physics and Astronomy Department, Johns Hopkins University, Baltimore, MD, 21218, \email{Shawn.Usman@jhu.edu}
\altaffiltext{2}{Space Telescope Science Institute, Baltimore, MD}
  21218} \altaffiltext{3}{Harvard-Smithsonian Center for Astrophysics,
  60 Garden Street, Cambridge, MA 02138}
\altaffiltext{3}{Department of Physics and Astronomy,
  Dartmouth College, 6127 Wilder Laboratory, Hanover, NH, 03755}
 \altaffiltext{4}{Department of Physics and Astronomy,
  University of Missouri, Kansas City, MO, 64110}

\begin{abstract}

 We explore the connection between absorption by neutral gas and
 extinction by dust in mid-infrared (IR) selected luminous quasars. We
 use a sample of 33 quasars at redshifts $0.7< z \lesssim 3$ in the 9
 deg$^2$ Bo\"{o}tes multiwavelength survey field that are selected
 using {\em Spitzer Space Telescope} Infrared Array Camera colors and
 are well-detected as luminous X-ray sources (with $>$150 counts) in
 {\em Chandra} observations. We divide the quasars into dust-obscured
 and unobscured samples based on their optical to mid-IR color, and
 measure the neutral hydrogen column density $N_{\rm H}$ through
 fitting of the X-ray spectra. We find that all subsets of quasars
 have consistent power law photon indices $\Gamma\approx1.9$ that are
 uncorrelated with $N_{\rm H}$. We classify the quasars as
 gas-absorbed or gas-unabsorbed if $N_{\rm H} > 10^{22} \;{\rm
   cm}^{-2}$ or $N_{\rm H} < 10^{22} \;{\rm cm}^{-2}$, respectively. Of
 24 dust-unobscured quasars in the sample, only one shows clear
 evidence for significant intrinsic $N_H$, while 22 have column
 densities consistent with $N_{\rm H} < 10^{22} \;{\rm cm}^{-2}$. In
 contrast, of the nine dust-obscured quasars, six show evidence for
 intrinsic gas absorption, and three are consistent with $N_{\rm H} <
 10^{22} \;{\rm cm}^{-2}$.  We conclude that dust extinction in
 IR-selected quasars is strongly correlated with significant gas
 absorption as determined through X-ray spectral fitting. These
 results suggest that obscuring gas and dust in quasars are generally co-spatial, and confirm the reliability of simple mid-IR and optical photometric techniques for separating quasars based on obscuration.

\end{abstract}

\keywords{galaxies: active -- infrared: galaxies -- quasars: general--
  surveys -- X-rays: galaxies}

\section{Introduction}

\defcitealias{hick07abs}{H07} \defcitealias{dono14qsoclust}{Donoso et al. 2014}

In the past decade, sensitive mid-infrared (IR) observations with the
{\em Spitzer Space Telescope} and {\em Wide-Field Infrared Explorer}
(WISE) have forever changed our understanding of BH growth by
unveiling, at last, large populations of obscured quasars. Many
luminous type 2 active galactic nuclei (AGNs) have been identified from
narrow optical emission lines (\citealt{zaka03,zaka04, zaka05,
  reye08qso2}), radio luminosity (e.g., \citealt{mcca93highzradio,
  mart06, seym07radiohosts}), or X-ray properties (e.g,
\citealt{alex01xfaint, ster02, trei04, vign06qso2, vign09qso2}), but
the most efficient techniques for finding obscured quasars employ
mid-IR photometry. Pioneering work with {\em Spitzer} showed that
obscured quasars have similar mid-IR SEDs to their unobscured
counterparts, but are dominated by host galaxy light in the optical
(\citealt{lacy04, ster05, rowa05, poll06, alon06,
  donl08spitz, hick07abs}, hereafter H07).   In the mid-IR, obscured quasars can be reliably
selected using simple color criteria, in particular being very red in
[3.6]--[4.5] (characteristic of a ``hot'' mid-IR SED) and very red in
$R-[4.5]$, indicating a faint optical counterpart
(Fig.1). Mid-IR studies find roughly equal numbers of
obscured and unobscured quasars (e.g., \citetalias{hick07abs, dono14qsoclust}).

Obscured quasars therefore represent a large fraction of the massive
BH growth in the Universe, but their precise nature remains a mystery.
In particular, what is the origin of the obscuring material, and what
role do these objects play in the evolution of black holes and
galaxies? The simple ``unified model'' for AGN attributes obscuration
to the orientation of a gas- and dust-rich torus intrinsic to the
central engine (e.g., \citealt{anto93, ball06b}), but it is not clear
whether this model applies to objects with quasar luminosities. A
competing hypothesis is that quasars are fueled by major mergers of
galaxies that drive gas and dust clouds to the nucleus, obscuring the
central engine, as suggested in models of BH-galaxy co-evolution
(e.g., \citealt{sand88,hopk08frame1}).

One clue about the nature of the obscuring material is the
connection between obscuration by {\em dust} which is manifested
through extinction of rest-frame ultraviolet and optical nuclear
light, and absorption by neutral {\em gas} which is detectable by its
effect on the observed X-ray spectrum. In the simplest unified
scenarios, neutral gas and dust are co-spatial and so both types of
obscuration should be observed in the same systems. This is in fact
what is observed in most obscured AGNs, however $\sim$30\% of
moderate-luminosity AGNs show a mismatch between optical and X-ray
classification criteria \citep[e.g.,][]{tozz06, trou09optx}, suggesting some
deviation from the simplest unified scenario. 

\begin{figure} 
 \centerline{\includegraphics[width=\columnwidth]{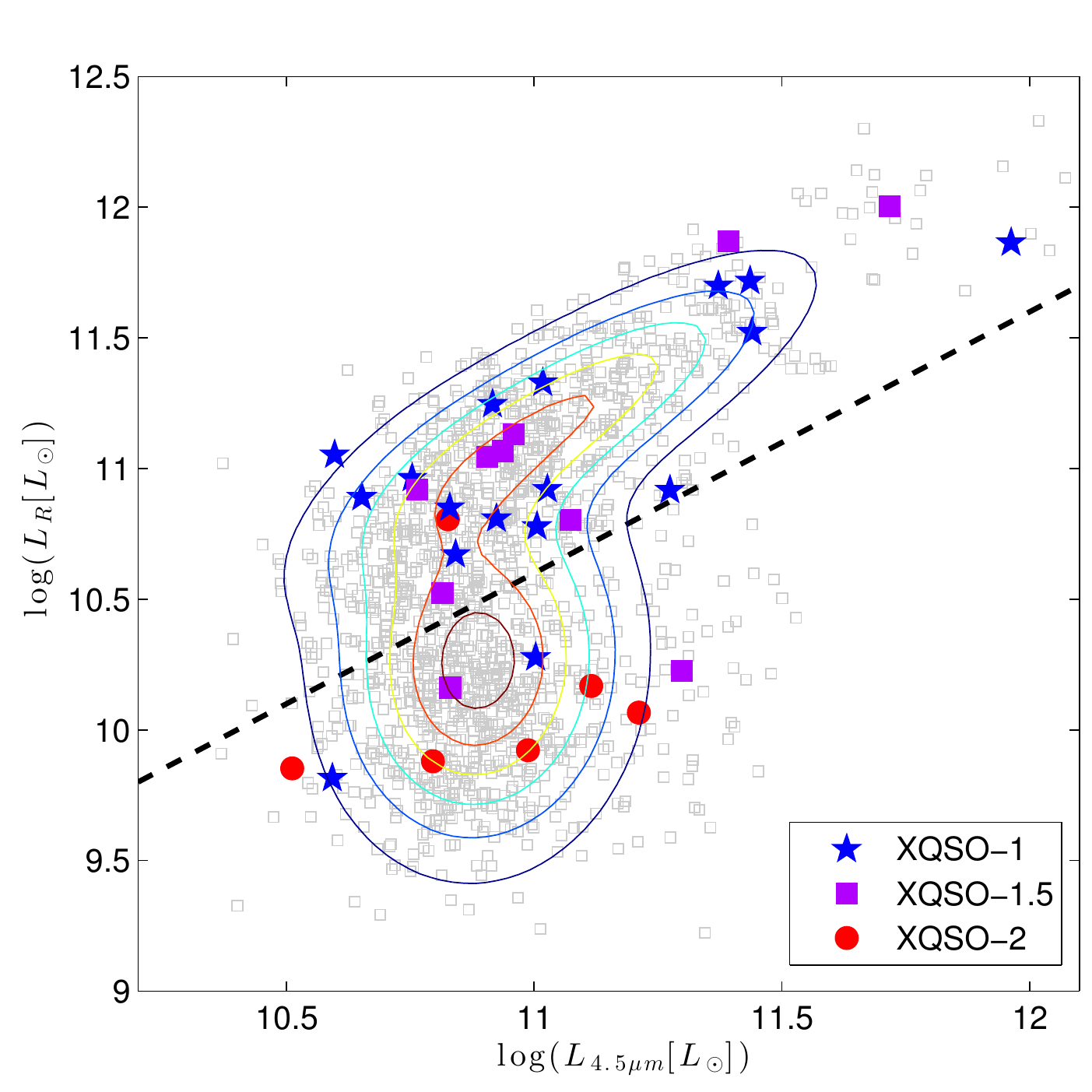}}
  \let\normalsize\footnotesize
  \caption{$L_{4.5 \mu m} vs. L_{R}$ (calculated as $\nu L_\nu$ in the
    observed frame) for IR-selected quasars. The selection boundary of
    $log(L_{R}/L_{4.5\mu})=-0.4$ (corresponding to $R-[4.5] = 6.1$ in
    Vega magnitudes) is shown as a dashed line with IRQSO-1s and
    IRQSO-2s residing above and below the line, respectively. Contours
    are derived from IR source density. The filled blue stars
    (XQSO-1), purple squares (XQSO-1.5), and red circles (XQSO-2) are
    the X-ray classifications for our X-ray spectroscopic sample,
    discussed in \S 3.\label{fig:1}}
\end{figure}

Obscuration is particularly interesting for powerful quasars (given the expected importance of merger fueling) but the obscuration properties of luminous AGN have been less well studied due to their relative rarity. Furthermore, it is also interesting to study the obscuration in luminous quasars given the
reported decreasing trend of obscured AGN at increasing luminosities (e.g., \citealt{ued03, has05, merl14agnobs}). A recent study of X-ray selected AGNs in the COSMOS field  \citet{merl14agnobs} found that at quasar luminosities ($L_X> 10^{44}$ erg s$^{-1}$) $\sim$80\% of AGNs have classifications for dust and gas obscuration that agree, with the majority of the "mismatch" corresponding to optically-unobscured but gas-obscured sources. Previous studies of such objects have suggested they are preferentially hosted by more rapidly star-forming galaxies, indicating that the X-ray absorbing gas arises from galaxy-scale structures that are far larger than the putative torus, and may be
associated with an outflowing wind \citep{page04submm}. However, \citet{merl14agnobs} 
found that  X-ray obscured, optically-unobscured quasars have spectra and photometric properties that are identical to their X-ray unobscured counterparts, suggesting that the absorption may instead be due to small-scale clouds within the putative torus. 

In this study, we further explore the connection between gas and dust obscuration in luminous quasars, using observations from the wide-field (9 deg$^2$) Bo\"{o}tes multiwavelength survey area that enable efficient selection of rare, luminous sources. We analyze X-ray spectra of 33 high-luminosity
IR-selected quasars in the XBo\"{o}tes Deep Survey.  We find a strong
correspondence between dust and gas obscuration in luminous quasars, consistent with the
predictions of unified models as well as simple evolutionary scenarios
in which the gas and dust are co-spatial.

\section{Observations and quasar sample}

Our sample of quasars is drawn from \citetalias{hick07abs}, who selected
1479 luminous AGNs based on the mid-IR color criteria of
\citet{ster05}. The vast majority of these sources have estimates of
bolometric luminosity $L_{\rm bol} > 10^{12} L_{\sun}$ corresponding
to the commonly-used division between ``Seyfert galaxies'' and
``quasars'' \citep[e.g.,][]{hopk09fueling}, so for the remainder of
this paper we refer to the sample objects as quasars. The mid-IR
observations come from the Spitzer IRAC Shallow Survey \citep{eise04,
  brod06} and optical photometry from the NOAO Deep Wide Field Survey
\citet{jann99}, and is limited to redshifts $0.7 < z < 3$ determined
by optical spectroscopy from the AGN and Galaxy Evolution Survey
(AGES; \citealt{koch12ages}) or using photometric redshifts from
\citet{brod06}. \citetalias{hick07abs} found that the full IR-selected
quasar sample could be easily divided into dust-obscured and
unobscured sources based on their optical to mid-IR color, as they
exhibited a bimodal color distribution with a boundary at $R-[4.5] = 6.1$
(or equivalently $\log(L_{R}/L_{{\rm 4.5\mu m}})=-0.4$, where $L_{R}$
and $L_{{\rm 4.5\mu m}}$ are the luminosities in the {\em
  observed}-frame $R$ and 4.5 micron bands; see
Figure~\ref{fig:1}). We also include estimates of quasar bolometric
luminosity derived from fitting mid-IR spectral energy distributions,
as described by \citet{hick11qsoclust}.

\citetalias{hick07abs} showed that the objects with blue $R-[4.5]$ colors
are dominated in the optical by unobscured light from the quasar
nucleus, while the nucleus is obscured in the redder sources such that
the optical light is dominated by the host galaxy.  \citetalias{hick07abs}
also showed via an X-ray stacking analysis that the average X-ray
spectra of unobscured quasars were consistent with no absorption by
neutral gas, while the obscured quasars had harder average X-ray
spectra indicating significant gas absorption with $N_{\rm H} \sim
3\times10^{22}$ cm$^{-2}$. However, given the shallow (5 ks;
\citealt{murr05}) X-ray observations available in the
\citetalias{hick07abs} analysis, this comparison was only possible for {\em
  average} X-ray hardness ratios and could not explore variations
in X-ray spectra between individual sources.

This study expands on that work by analyzing deeper {\em Chandra} ACIS
observations (including from the XBo\"{o}tes Deep Survey) of quasars
in the \citetalias{hick07abs} sample.  The deeper observations consist of
36 exposures between 2001 and 2012 with an average exposure time of
41.6 ks, of which 13 are with the ACIS-I array and 13 are with ACIS-S
(Table \ref{tab:1}). The data were reprocessed using the CHAV
(v.2011.01.25) and CIAO 4.1.2 (CALDB 4.1.4) packages.  To obtain
sufficient quality X-ray spectra we limit our analyses to objects with
$>$150 counts in the 0.2-7 keV band. We match the sources to the 1479
IR-selected quasars from \citetalias{hick07abs} using the TOPCAT
\citep{tayl05topcat} cone-search algorithm with a 10\arcsec\ radius. A
total of 33 quasars were selected, with an average of $\approx\;300$ counts
per source.

Using the \citetalias{hick07abs} $R-[4.5]$ color criterion, 24 of the
IR-selected quasars in our sample are classified as unobscured
(IRQSO-1) while 9 are obscured (IRQSO-2).  Our bright X-ray
spectroscopic sample contains a low fraction of IRQSO-2s (9/33 or
27\%) that is significantly smaller than the 43\% obscured fraction in
the full IRQSO sample from \citetalias{hick07abs}. This is due to the fact that X-ray QSO2s are
generally fainter in X-rays; the fraction of IRQSOs with X-ray
counterparts in original shallow (5 ks) XBo\"{o}tes survey is
$\approx$60\% for IRQSO-1s and only $\approx$30\% for IRQSO-2s, while
the average X-ray flux of the undetected sources, determined from
stacking, is also systematically lower for the IRQSO-2s
\citepalias{hick07abs}. The fainter X-ray emission from IRQSO-2s is most
likely due to higher levels of gas obscuration that not only hardens
the X-ray spectrum but reduces the observed flux
\citep[e.g.,][]{alex08compthick}.  This fact reflects a correspondence
between gas and dust obscuration in these quasars, which we aim to
test further by direct measurement of the obscuring column density
($N_{\rm H}$) in the X-ray spectra of the bright sources for which
these measurements are possible.

\begin{figure*}[t]
 \centerline{\includegraphics[width=7.0in]{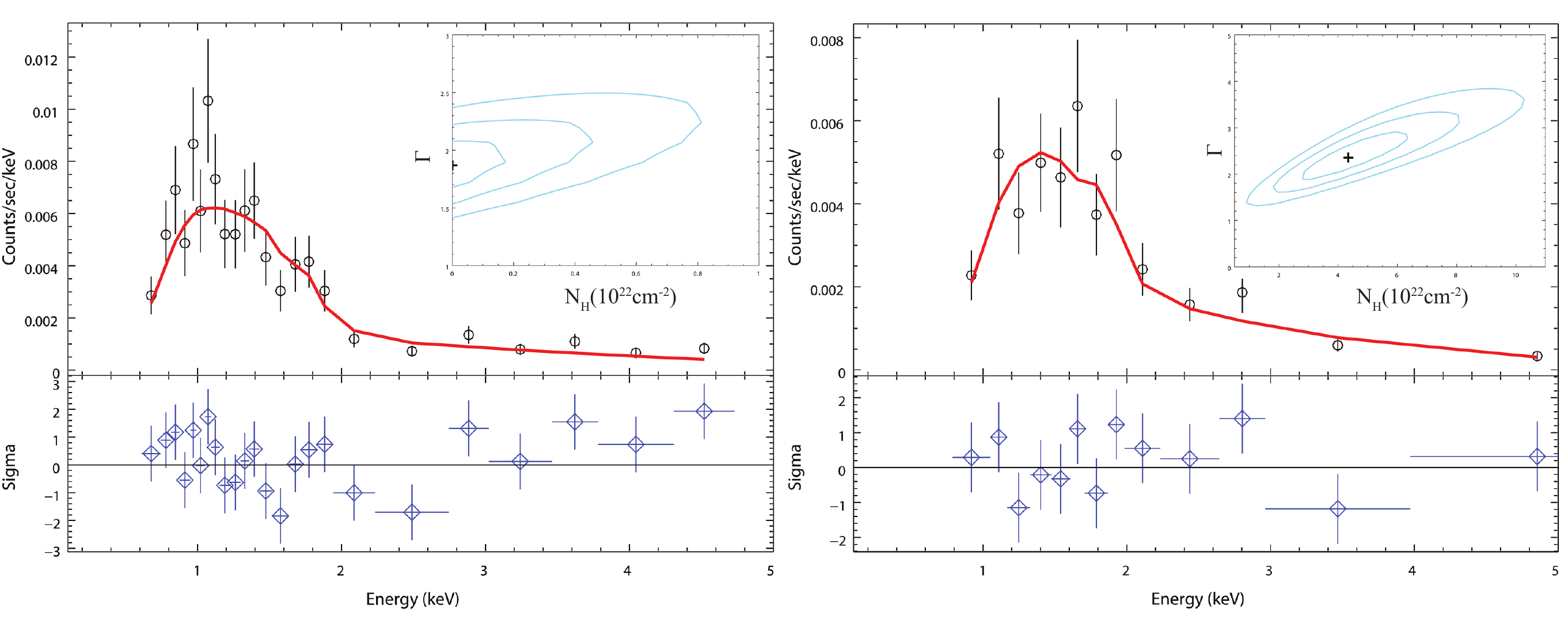}}
  \let\normalsize\footnotesize
\caption{(a) X-ray spectrum with best-fit model an gas-unobscured XQSO-1. Contours of $N_{\rm H}$ vs. $\Gamma$ are shown in the subset.  (b)
  Same as (a), but for an gas-obscured XQSO-2. A clear cut-off in
  spectra for energies $<$ 1 keV is absent in the XQSO-1 (a), but
  observed in the XQSO-2 (b).  
\label{fig:2}}
\end{figure*}

\section{X-Ray Analysis}

 We extract source and background spectra and produce response
 functions using standard CIAO software. Sources and background spectra were extracted from hand-selected circular regions with mean radii of 11.5 pixels and 100.1 pixels, respectively. The resulting spectra were
 fit using SHERPA (4.1.2) with a one-dimensional power law
 (xspowerlaw.p1) convolved with two neutral hydrogen absorption laws,
 one for the Milky Way (xswabs.abs1), and one for the host galaxy
 (xszwabs.abs2) absorption. The Milky Way absorption column density
 was frozen at the mean of all 33 quasars, $N_{\rm H-MW} = 1.05 \times
 10^{20}$ cm$^{-2}$, derived from the Chandra X-ray Center's COLDEN
 calculator. The host galaxy redshifts are fixed at values taken
 either from AGES spectroscopic measurements \citep{koch12ages} or
 photometric redshifts \citep{brod06}, as described in \citetalias{hick07abs}. The power-law index $\Gamma$, normalization, and
 intrinsic $N_{\rm H}$ were allowed to vary in the spectral
 fitting. Examples of the X-ray spectral fits are shown in Figure
 \ref{fig:2} for one source showing no evidence for gas absorption, and another with detectable $N_{\rm H}$.

\begin{deluxetable*}{ccccccccc}
\tabletypesize{\scriptsize}
\tablecolumns{9} 
\tablewidth{0pt}
\setlength{\tabcolsep}{0.02in} 
\tablecaption{Spectral Parameters and X-ray Classification \label{tab:1}}
\tablenum{1}
\tablehead{\colhead{Identifier} & \colhead{Exposure} &
  \colhead{z\tablenotemark{a}} & \colhead{Counts} & \colhead{$F_{x}$ \tablenotemark{b}} & \colhead{
log$(L_{\rm x})$} & \colhead{$\Gamma$} & \colhead{$N_{\rm H}\pm 1\sigma\tablenotemark{c}$}
  & \colhead{XQSO}\\
 &(ks) & & (0.5-7 keV) & $\mathrm{(10^{-14} ergs \ cm^{-2} \ s^{-1})}$ & $\mathrm{(ergs \ s^{-1})} $ & & ($10^{22}$  cm$^{-2}$) & Class} 

\startdata
\cutinhead{IRQSO-1}
SDWFS J142942.63+335654.94 & 47.1 & 1.1251 & 607 & 17.3 & 45.1 & $1.77 \pm 0.09$ & $0+0.07$ & 1 \\
SDWFS J142810.31+353847.31 & 37.9 & 0.8028 & 852 & 16.5 & 44.7 & $1.80 \pm 0.07$ & $0+0.02$ & 1 \\
SDWFS J143520.20+340929.20 & 42.6 & 1.0972 & 182 & 1.75 & 44.1 & $2.13 \pm 0.20$ & $0+0.08$ & 1 \\
SDWFS J143520.60+340514.68 & 42.6 & 0.7973 & 217 & 3.38 & 44.0 & $1.59 \pm 0.14$ & $0+0.06$ & 1 \\
SDWFS J143513.41+350053.77 & 44.0 & 1.1471 & 325 & 3.50 & 44.4 & $1.81 \pm 0.13$ & $0+0.15$ & 1 \\
SDWFS J143520.14+350413.23 & 44.0 & 1.0512 & 339 & 1.10 & 43.8 & $2.30 \pm 0.18$ & $0+0.13$ & 1 \\
SDWFS J142922.94+351517.74 & 42.0 & 0.9041 & 408 & 10.4 & 44.6 & $1.87 \pm 0.14$ & $0+0.09$ & 1 \\
SDWFS J142634.05+351602.56 & 14.9 & 1.1055 & 177 & 6.61 & 44.7 & $1.61 \pm 0.17$ & $0+0.17$ & 1 \\
SDWFS J143651.98+350537.97 & 54.3 & 0.864   & 200 & 0.54 & 43.3 & $2.49 \pm 0.21$ & $0+0.05$ & 1 \\
SDWFS J142651.47+351924.40 & 29.7 & 1.756   & 833 & 12.4 & 45.4 & $1.83 \pm 0.07$ & $0+0.07$ & 1 \\ 
SDWFS J142839.20+353455.55 & 37.9 & 1.0693 & 228 & 5.31 & 44.5 & $1.70 \pm 0.16$ & $0+0.35$ & 1 \\
SDWFS J142829.91+342758.68 & 68.6 & 1.1401 & 301 & 3.31 & 44.4 & $1.93 \pm 0.16$ & $0+0.41$ & 1 \\
SDWFS J143245.88+333758.45 & 33.8 & 1.0816 & 192 & 3.43 & 44.3 & $2.13 \pm 0.26$ & $0+0.34$ & 1 \\
SDWFS J142917.20+342130.31 & 28.7 & 1.2734 & 306 & 6.06 & 44.8 & $2.07 \pm 0.18$ & $0+0.56$ & 1 \\
SDWFS J142607.71+340426.61 & 51.2 &  4.32    & 468 & 7.27 & 46.1 & $2.67 \pm 0.12$ & $0+0.94$ & 1 \\
SDWFS J143310.25+335421.99 & 27.7 & 0.8792 & 238 & 9.62 & 44.6 & $1.79 \pm 0.33$ & $0.57+0.75$ & 1.5 \\
SDWFS J143153.67+344138.17 & 50.2 & 0.8254 & 481 & 8.16 & 44.4 & $1.72 \pm 0.17$ & $1.31\pm 0.30$ & 2 \\
SDWFS J142658.70+324003.78 & 42.0 & 1.7737 & 212 & 5.45 & 45.1 & $1.56 \pm 0.25$ & $0+1.67$ & 1.5 \\
SDWFS J142915.19+343820.18 & 29.7 & 2.3515 & 202 & 3.87 & 45.2 & $2.12 \pm 0.27$ & $0+1.83$ & 1.5 \\
SDWFS J142622.68+334202.41 & 33.8 & 1.3517 & 182 & 4.27 & 44.7 & $1.94 \pm 0.31$ & $0.42+1.41$ & 1.5 \\
SDWFS J143434.40+330549.14 & 56.8 & 2.0656 & 154 & 0.49 & 44.2 & $2.39 \pm 0.45$ & $0.76+2.38$ &  1.5 \\
SDWFS J142859.55+350349.11 & 42.0 & 1.6126 & 165 & 2.89 & 44.7 & $2.19 \pm 0.45$ & $1.28+2.21$ &  1.5 \\
SDWFS J143431.07+332825.51 & 39.9 & 1.0554 & 155 & 4.70 & 44.5 & $1.76 \pm 0.38$ & $1.45+1.68$ & 1.5 \\
SDWFS J143450.01+352520.65 & 30.7 & 0.9629 & 184 & 8.17 & 44.6 & $2.56 \pm 0.42$ & $1.98\pm 1.09$ & 1.5 \\
\cutinhead{IRQSO-2}
SDWFS J143210.97+343957.89 & 81.9 & 0.83 & 204 & 1.93 & 43.8 & $1.66 \pm 0.24$ & $0+0.66$& 1       \\
SDWFS J142824.97+352842.63 & 37.9 & 1.49 & 217 & 4.90 & 44.8 & $1.98 \pm 0.20$ & $0+0.71$&  1\\
SDWFS J142845.07+350903.30 & 42.0 & 0.81 & 463 & 9.52 & 44.5 & $1.87 \pm 0.20$ & $0.72 \pm 0.42$ & 1.5\\
SDWFS J142813.98+325502.82 & 33.8 & 1.54 & 264 & 10.8 & 45.2 & $1.08 \pm 0.28$ & $2.21+3.35$& 1.5  \\
SDWFS J143502.04+330556.51 & 56.8 & 1.58 & 299 & 5.09 & 44.9 & $1.84 \pm 0.25$ & $3.07 \pm 1.62$&   2\\
SDWFS J143503.49+340241.92 & 42.6 & 1.06 & 219 & 2.54 & 44.2 & $2.85 \pm 0.44$ & $3.41 \pm 1.35$&  2 \\
SDWFS J143359.09+331301.06 & 42.0 & 0.92 & 330 & 17.3 & 44.9 & $1.64 \pm 0.31$ & $3.97 \pm 1.49$&  2 \\
SDWFS J142916.10+335537.36 & 24.6 & 0.98 & 220 & 7.30 & 44.6 & $2.36 \pm 0.34$ & $4.34 \pm 1.21$ & 2\\ 
SDWFS J142707.05+325214.17 & 42.0 & 2.28 & 244 & 5.65 & 45.4 & $1.97 \pm 0.28$ & $5.66 \pm 2.98$&  2
\enddata

\tablenotetext{a}{Differences in redshift precision correspond to the
 differences between spectroscopic and photometric estimates ($\S2$).}
\tablenotetext{b}{Fluxes and luminosities are unabsorbed and calculated in the 0.5-7keV band.}
\tablenotetext{c}{Entries with plus signs only indicate that the lower bound is equal to zero.}

\end{deluxetable*}

\begin{figure}[!h] 
 \centerline{\includegraphics[width=\columnwidth]{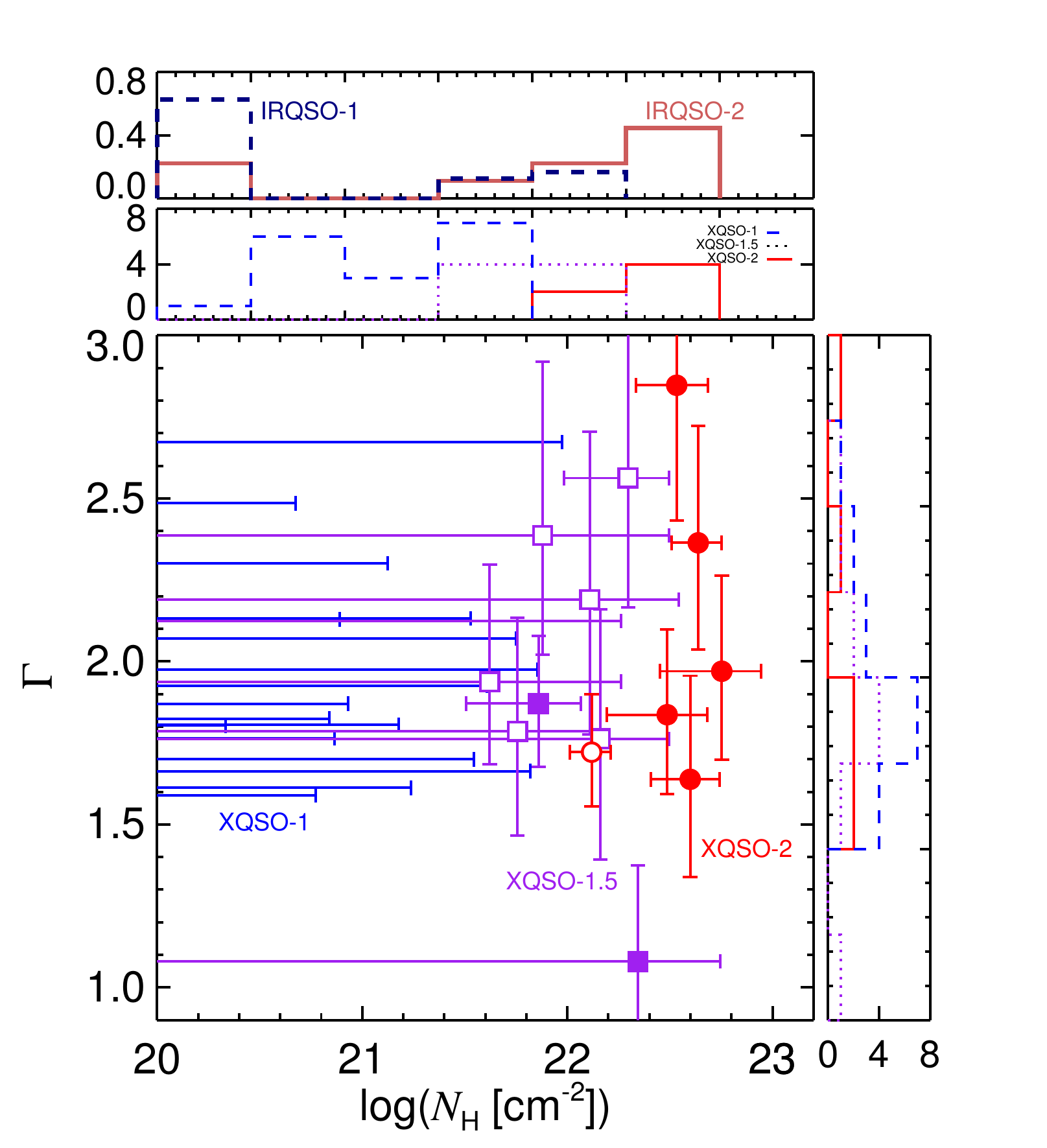}}
  \let\normalsize\footnotesize
  \caption{X-ray spectral fit parameters for the full sample of 33 IR-
    and X-ray selected quasars. The main panel shows $\Gamma$ and
    $N_{\rm H}$ with uncertainties, highlighting the populaton of
    XQSO-1.5s with ambiguous classifications, and showing that there
    is no clear correlation between $\Gamma$ and $N_{\rm H}$. Dust unobscured (IRQSO-1) sources are marked with open symbols. The top
    panel shows the fractional distribution of the best-fit values of
    $N_{\rm H}$ for the IRQSO-1s and 2s, respectively, showing the
    clear correlation between dust and gas obscuration. The other two
    panels show the distributions in best-fit $N_{\rm H}$ and $\Gamma$
    for the XQSO-1s, 1.5s, and 2s separately.
\label{fig:gamma_nh}}
\end{figure}

The spectral fit parameters ($\Gamma$ and intrinsic $N_{\rm H}$) and $1\sigma$ errors along
with other quasar properties are shown in Table 1. In many cases, the observed $N_{\rm H}$ is
consistent with zero and only upper limits are obtained. We classify the quasars as gas-absorbed (XQSO-2) or gas-unabsorbed (XQSO-1) if $N_{\rm H} > 10^{22} \;{\rm cm}^{-2}$ or $N_{\rm H} < 10^{22} \;{\rm
  cm}^{-2}$, respectively, following the convention used widely in the literature (e.g., \citealt{tozz06, lan13, ued14}). For several sources (XQSO-1.5) the classification is ambiguous, with $N_{\rm H}$ consistent with either
subset within the uncertainties. (We note that a division at $N_{\rm H}=10^{21.5}$ cm$^{-2}$, as in \citealt{merl14agnobs}, produces more ambiguous classifications due to the uncertainties in our $N_{\rm H}$ measurements, but results in the same qualitative conclusions.) In Figure \ref{fig:1} we show the IR
and optical luminosities of the quasars in different X-ray classes,
and in Figure \ref{fig:gamma_nh} we show the distribution in $\Gamma$
and $N_{\rm H}$ for all the sources. We note that there is no clear
correlation between $\Gamma$ and $N_{\rm H}$, verifying that we have
sufficient counts in each source to sufficiently break the degeneracy
between those parameters. The error-weighted average value of $\Gamma$
is $1.89\pm0.03$ for the full sample, and for the subsets is
$1.90\pm0.03$ (XQSO-1), $1.83\pm0.10$ (XQSO-1.5) and $1.90\pm0.11$
(XQSO-2), all consistent with previous measurements for the intrinsic
photon index for AGNs and quasars \citep[e.g.,][]{tozz06, xue11cdfs}.

The primary result of this analysis is the correspondence between the
optical/mid-IR and X-ray obscuration criteria. Among the 24
dust-unobscured (IRQSO-1) quasars, we find 15 XQSO-1s, 8 XQSO-1.5s (of
which 7 have $N_{\rm H}$ consistent with zero), and only one
XQSO-2. In contrast, the 9 dust-obscured quasars (IRQSO-2s) comprise
only 2 XQSO-1s, 2 XQSO-1.5s (of which one has $N_{\rm H}$ consistent
with zero) and 5 XQSO 2s. We thus obtain a strong, if not perfect,
correlation between absorption by gas and dust, with only a few
examples showing clearly anomalous gas absorption for the observed
optical and mid-IR properties (broadly consistent with the results of \citealt{merl14agnobs}. The largest $N_{\rm
  H}$ observed in our XQSO-2 sample is relatively modest (only a few
$\times10^{22}$ cm$^{-2}$); it is likely that more heavily obscured
sources ($N_{\rm H}>10^{23}$) are excluded from our X-ray
spectroscopic sample because their observed fluxes are too faint to
yield the required numbers of counts, as discussed in \S~2.

\section{Discussion and conclusions}

Overall, these results indicate that obscuration by gas and dust are
strongly correlated in IR-selected luminous quasars. This
correspondence indicates that gas and dust obscuration generally arise
in the same structures, consistent with the predictions of the unified
model but also with the simplest evolutionary models. We thus cannot
draw robust conclusions about whether the obscuring material is in a
small-scale torus or due to larger-scale galactic structures.  A
recent study of their far-IR properties using 250 $\mu$m data from
{\em Herschel} suggests that the IRQSO-2s have higher average rates of
star formation than IRQSO-1s (Chen et al.\ 2014, in prep), suggesting
that at least some quasars are obscured by galaxy-scale material
associated with rapid star formation. However, only 4 of our 33
objects are detected at 250 $\mu$m with {\em Herschel} and of these
two show neither gas nor dust obscuration, while the other two are
obscured in both classifications, so it is unclear whether large-scale
material is responsible for any of the observed obscuration.

We find that only a small fraction of dust-unobscured quasars show
{\em clear} X-ray absorption signatures. Of the quasars in our X-ray
spectroscopic sample, $\approx$17\% are IRQSO-1s with best-fit values
of $N_{\rm H}>10^{22}$ cm$^{-2}$, similar to the fraction of such
mismatches found by \citet{merl14agnobs}. We note however that all but
one of these objects are consistent with having little or no gas
absorption, and their distribution in the plane IR and optical
luminosity (Figure 1) is indistinguishable to IRQSO-1s with no X-ray
absorption. We therefore caution that the observed incidence of these
mismatches must be treated as an upper limit, and that the X-ray
absorbed broad-line quasars may in fact be rare in the full quasar
population; an accurate census of quasars with modest obscuration
provides strong motivation for future high-throughput X-ray
observatories.

Overall our results serve to verify that obscured quasar selection based on
optical to mid-IR color \citetalias{hick07abs} preferentially identifies
systems that show evidence for obscuration at other wavelengths. Even
in our bright X-ray spectroscopic sample, for which the effective flux
limit biases us {\em against} heavily X-ray obscured sources as
discussed in \S~2, we still find that the majority of dust-obscured
quasar candidates show clear evidence for X-ray absorption. This indicates that the full population of dust-obscured quasars likely has a very high incidence of corresponding gas absorption. These results 
confirm that optical/mid-IR color selection is effective in
selecting even moderately obscured quasars at the highest
luminosities, providing a strong basis for future large statistical studies of obscured quasars selected based on WISE and optical photometry.

\acknowledgments

This work was supported by NASA through ADAP award NNX12AE38G and by the National Science Foundation through grant number 1211096.This work is based on observations made with the Spitzer Space Telescope, which is operated by the Jet Propulsion Laboratory, California Institute of Technology under a contract with NASA. This work is based on observations with the Chandra X-ray Telescope, which is operated by SAO under a contract with NASA NAS8-03060.

{\it Facilities:} \facility{CXO (ACIS)}, \facility{Spitzer}, \facility{Mayall (MOSAIC-1)}


\begin{thebibliography}{38}
\expandafter\ifx\csname natexlab\endcsname\relax\def\natexlab#1{#1}\fi

\bibitem[{{Alexander} {et~al.}(2001){Alexander}, {Brandt}, {Hornschemeier},
  {Garmire}, {Schneider}, {Bauer}, \& {Griffiths}}]{alex01xfaint}
{Alexander}, D.~M., {Brandt}, W.~N., {Hornschemeier}, A.~E., {Garmire}, G.~P.,
  {Schneider}, D.~P., {Bauer}, F.~E., \& {Griffiths}, R.~E. 2001, \aj, 122,
  2156

\bibitem[{{Alexander} {et~al.}(2008){Alexander}, {Chary}, {Pope}, {Bauer},
  {Brandt}, {Daddi}, {Dickinson}, {Elbaz}, \& {Reddy}}]{alex08compthick}
{Alexander}, D.~M., et~al.\ 2008,  \apj, 687, 835

\bibitem[{{Alonso-Herrero} {et~al.}(2006){Alonso-Herrero}, {P{\'e}rez-Gonz
  {\'a}lez}, {Alexander}, {Rieke}, {Rigopoulou}, {Le Floc'h}, {Barmby},
  {Papovich}, {Rigby}, {Bauer}, {Brandt}, {Egami}, {Willner}, {Dole}, \&
  {Huang}}]{alon06}
{Alonso-Herrero}, A., et~al.\ 2006, \apj, 640, 167

\bibitem[{{Antonucci}(1993)}]{anto93}
{Antonucci}, R. 1993, \araa, 31, 473

\bibitem[{{Ballantyne} {et~al.}(2006){Ballantyne}, {Shi}, {Rieke}, {Donley},
  {Papovich}, \& {Rigby}}]{ball06b}
{Ballantyne}, D.~R., {Shi}, Y., {Rieke}, G.~H., {Donley}, J.~L., {Papovich},
  C., \& {Rigby}, J.~R. 2006, \apj, 653, 1070

\bibitem[{{Brodwin} {et~al.}(2006){Brodwin}, {Brown}, {Ashby}, {Bian}, {Brand},
  {Dey}, {Eisenhardt}, {Eisenstein}, {Gonzalez}, {Huang}, {Jannuzi},
  {Kochanek}, {McKenzie}, {Murray}, {Pahre}, {Smith}, {Soifer}, {Stanford},
  {Stern}, \& {Elston}}]{brod06}
{Brodwin}, M., et~al.\ 2006, \apj, 651, 791

\bibitem[{{Donley} {et~al.}(2008){Donley}, {Rieke}, {P{\'e}rez-Gonz{\'a}lez},
  \& {Barro}}]{donl08spitz}
{Donley}, J.~L., {Rieke}, G.~H., {P{\'e}rez-Gonz{\'a}lez}, P.~G., \& {Barro},
  G. 2008, \apj, 687, 111

\bibitem[{{Donoso} {et~al.}(2014){Donoso}, {Yan}, {Stern}, \&
  {Assef}}]{dono14qsoclust}
{Donoso}, E., {Yan}, L., {Stern}, D., \& {Assef}, R.~J. 2014, \apj\ submitted
  (arXiv:1309.2277)

\bibitem[{{Eisenhardt} {et~al.}(2004){Eisenhardt}, {Stern}, {Brodwin}, {Fazio},
  {Rieke}, {Rieke}, {Werner}, {Wright}, {Allen}, {Arendt}, {Ashby}, {Barmby},
  {Forrest}, {Hora}, {Huang}, {Huchra}, {Pahre}, {Pipher}, {Reach}, {Smith},
  {Stauffer}, {Wang}, {Willner}, {Brown}, {Dey}, {Jannuzi}, \&
  {Tiede}}]{eise04}
{Eisenhardt}, P.~R., et~al.\ 2004, \apjs, 154, 48

\bibitem[{{Hasinger} {et~al.}(2005){Hasinger}, {Miyaji}, \& {Schmidt}}]{has05}
{Hasinger}, G.,{Miyaji}, T., \& {Schmidt}, M. \ 2005, \aap, 441, 417

\bibitem[{{Hickox} {et~al.}(2007){Hickox}, {Jones}, {Forman}, {Murray},
  {Brodwin}, {Brown}, {Eisenhardt}, {Stern}, {Kochanek}, {Eisenstein}, {Cool},
  {Jannuzi}, {Dey}, {Brand}, {Gorjian}, \& {Caldwell}}]{hick07abs}
{Hickox}, R.~C., et~al.\ 2007, \apj, 671, 1365

\bibitem[{{Hickox} {et~al.}(2011){Hickox}, {Myers}, {Brodwin}, {Alexander},
  {Forman}, {Jones}, {Murray}, {Brown}, {Cool}, {Kochanek}, {Dey}, {Jannuzi},
  {Eisenstein}, {Assef}, {Eisenhardt}, {Gorjian}, {Stern}, {Le Floc'h},
  {Caldwell}, {Goulding}, \& {Mullaney}}]{hick11qsoclust}
{Hickox}, R.~C., et~al.\ 2011, \apj, 731, 117

\bibitem[{{Hopkins} \& {Hernquist}(2009)}]{hopk09fueling}
{Hopkins}, P.~F. \& {Hernquist}, L. 2009, \apj, 694, 599

\bibitem[{{Hopkins} {et~al.}(2008){Hopkins}, {Hernquist}, {Cox}, \& {Kere{\v
  s}}}]{hopk08frame1}
{Hopkins}, P.~F., {Hernquist}, L., {Cox}, T.~J., \& {Kere{\v s}}, D. 2008,
  \apjs, 175, 356

\bibitem[{{Jannuzi} \& {Dey}(1999)}]{jann99}
{Jannuzi}, B.~T. \& {Dey}, A. 1999, in ASP Conf.\ Ser.\ 191: Photometric
  Redshifts and the Detection of High Redshift Galaxies, ed. R.~{Weymann},
  L.~{Storrie-Lombardi}, M.~{Sawicki}, \& R.~{Brunner} (San Francisco: ASP),
  111

\bibitem[{{Kochanek} {et~al.}(2012){Kochanek}, {Eisenstein}, {Cool},
  {Caldwell}, {Assef}, {Jannuzi}, {Jones}, {Murray}, {Forman}, {Dey}, {Brown},
  {Eisenhardt}, {Gonzalez}, {Green}, \& {Stern}}]{koch12ages}
{Kochanek}, C.~S., et~al.\ 2012, \apjs, 200, 8

\bibitem[{{Lacy} {et~al.}(2004){Lacy}, {Storrie-Lombardi}, {Sajina},
  {Appleton}, {Armus}, {Chapman}, {Choi}, {Fadda}, {Fang}, {Frayer},
  {Heinrichsen}, {Helou}, {Im}, {Marleau}, {Masci}, {Shupe}, {Soifer},
  {Surace}, {Teplitz}, {Wilson}, \& {Yan}}]{lacy04}
{Lacy}, M., et~al.\ 2004, \apjs, 154, 166

\bibitem[{{Lanzuisi} {et~al.}(2013){Lanzuisi}, {Civano}, {Elvis}, {Salvato},
  {Hasinger}, {Vignali}, {Zamorani}, {Aldcroft}, {Brusai}, {Comastri},
  {Fiore}, {Fruscione}, {Gilli}, {Ho}, {Mainieri}, {Merloni}, \& {Zamorani}}]{lan13}
{Lanzuisi}, A., et~al.\ 2013, \mnras, 431, 978

\bibitem[{{Mart{\'{\i}}nez-Sansigre} {et~al.}(2006){Mart{\'{\i}}nez-Sansigre},
  {Rawlings}, {Lacy}, {Fadda}, {Jarvis}, {Marleau}, {Simpson}, \&
  {Willott}}]{mart06}
{Mart{\'{\i}}nez-Sansigre}, A., {Rawlings}, S., {Lacy}, M., {Fadda}, D.,
  {Jarvis}, M.~J., {Marleau}, F.~R., {Simpson}, C., \& {Willott}, C.~J. 2006,
  \mnras, 370, 1479

\bibitem[{{McCarthy}(1993)}]{mcca93highzradio}
{McCarthy}, P.~J. 1993, \araa, 31, 639

\bibitem[{{Merloni} {et~al.}(2014){Merloni}, {Bongiorno}, {Brusa}, {Iwasawa},
  {Mainieri}, {Magnelli}, {Salvato}, {Berta}, {Cappelluti}, {Comastri},
  {Fiore}, {Gilli}, {Koekemoer}, {Le Floc'h}, {Lusso}, {Lutz}, {Miyaji},
  {Pozzi}, {Riguccini}, {Rosario}, {Silverman}, {Symeonidis}, {Treister},
  {Vignali}, \& {Zamorani}}]{merl14agnobs}
{Merloni}, A., et~al.\ 2014, \mnras, 437, 3550

\bibitem[{{Murray} {et~al.}(2005){Murray}, {Kenter}, {Forman}, {Jones},
  {Green}, {Kochanek}, {Vikhlinin}, {Fabricant}, {Fazio}, {Brand}, {Brown},
  {Dey}, {Jannuzi}, {Najita}, {McNamara}, {Shields}, \& {Rieke}}]{murr05}
{Murray}, S.~S., et~al.\ 2005, \apjs, 161, 1

\bibitem[{{Page} {et~al.}(2004){Page}, {Stevens}, {Ivison}, \&
  {Carrera}}]{page04submm}
{Page}, M.~J., {Stevens}, J.~A., {Ivison}, R.~J., \& {Carrera}, F.~J. 2004,
  \apjl, 611, L85

\bibitem[{{Polletta} {et~al.}(2006){Polletta}, {Wilkes}, {Siana}, {Lonsdale},
  {Kilgard}, {Smith}, {Kim}, {Owen}, {Efstathiou}, {Jarrett}, {Stacey},
  {Franceschini}, {Rowan-Robinson}, {Babbedge}, {Berta}, {Fang}, {Farrah},
  {Gonz{\'a}lez-Solares}, {Morrison}, {Surace}, \& {Shupe}}]{poll06}
{Polletta}, M.~d.~C., et~al.\ 2006, \apj, 642, 673

\bibitem[{{Reyes} {et~al.}(2008){Reyes}, {Zakamska}, {Strauss}, {Green},
  {Krolik}, {Shen}, {Richards}, {Anderson}, \& {Schneider}}]{reye08qso2}
{Reyes}, R., et~al.\ 2008,  \aj, 136, 2373

\bibitem[{{Rowan-Robinson} {et~al.}(2005){Rowan-Robinson}, {Babbedge},
  {Surace}, {Shupe}, {Fang}, {Lonsdale}, {Smith}, {Polletta}, {Siana},
  {Gonzalez-Solares}, {Xu}, {Owen}, {Davoodi}, {Dole}, {Domingue},
  {Efstathiou}, {Farrah}, {Fox}, {Franceschini}, {Frayer}, {Hatziminaoglou},
  {Masci}, {Morrison}, {Nandra}, {Oliver}, {Onyett}, {Padgett},
  {Perez-Fournon}, {Serjeant}, {Stacey}, \& {Vaccari}}]{rowa05}
{Rowan-Robinson}, M., et~al.\ 2005, \aj, 129, 1183

\bibitem[{{Sanders} {et~al.}(1988){Sanders}, {Soifer}, {Elias}, {Madore},
  {Matthews}, {Neugebauer}, \& {Scoville}}]{sand88}
{Sanders}, D.~B., {Soifer}, B.~T., {Elias}, J.~H., {Madore}, B.~F., {Matthews},
  K., {Neugebauer}, G., \& {Scoville}, N.~Z. 1988, \apj, 325, 74

\bibitem[{{Seymour} {et~al.}(2007){Seymour}, {Stern}, {De Breuck}, {Vernet},
  {Rettura}, {Dickinson}, {Dey}, {Eisenhardt}, {Fosbury}, {Lacy}, {McCarthy},
  {Miley}, {Rocca-Volmerange}, {R{\"o}ttgering}, {Stanford}, {Teplitz}, {van
  Breugel}, \& {Zirm}}]{seym07radiohosts}
{Seymour}, N., et~al.\ 2007,  \apjs, 171, 353

\bibitem[{{Stern} {et~al.}(2005){Stern}, {Eisenhardt}, {Gorjian}, {Kochanek},
  {Caldwell}, {Eisenstein}, {Brodwin}, {Brown}, {Cool}, {Dey}, {Green},
  {Jannuzi}, {Murray}, {Pahre}, \& {Willner}}]{ster05}
{Stern}, D., et~al.\ 2005, \apj, 631, 163

\bibitem[{{Stern} {et~al.}(2002){Stern}, {Moran}, {Coil}, {Connolly}, {Davis},
  {Dawson}, {Dey}, {Eisenhardt}, {Elston}, {Graham}, {Harrison}, {Helfand},
  {Holden}, {Mao}, {Rosati}, {Spinrad}, {Stanford}, {Tozzi}, \& {Wu}}]{ster02}
{Stern}, D., et~al.\ 2002, \apj,  568, 71

\bibitem[{{Taylor}(2005)}]{tayl05topcat}
{Taylor}, M.~B. 2005, in Astronomical Society of the Pacific Conference Series,
  Vol. 347, Astronomical Data Analysis Software and Systems XIV, ed.
  P.~{Shopbell}, M.~{Britton}, \& R.~{Ebert}, 29

\bibitem[{{Tozzi} {et~al.}(2006){Tozzi}, {Gilli}, {Mainieri}, {Norman},
  {Risaliti}, {Rosati}, {Bergeron}, {Borgani}, {Giacconi}, {Hasinger},
  {Nonino}, {Streblyanska}, {Szokoly}, {Wang}, \& {Zheng}}]{tozz06}
{Tozzi}, P., et~al.\ 2006, \aap, 451, 457

\bibitem[{{Treister} {et~al.}(2004){Treister}, {Urry}, {Chatzichristou},
  {Bauer}, {Alexander}, {Koekemoer}, {Van Duyne}, {Brandt}, {Bergeron},
  {Stern}, {Moustakas}, {Chary}, {Conselice}, {Cristiani}, \&
  {Grogin}}]{trei04}
{Treister}, E., et~al.\ 2004, \apj, 616, 123

\bibitem[{{Trouille} {et~al.}(2009){Trouille}, {Barger}, {Cowie}, {Yang}, \&
  {Mushotzky}}]{trou09optx}
{Trouille}, L., {Barger}, A.~J., {Cowie}, L.~L., {Yang}, Y., \& {Mushotzky},
  R.~F. 2009, \apj, 703, 2160

\bibitem[{{Ueda} {et~al.}(2014){Ueda}, {Akiyama}, {Hasinger}, {Miyaji}, \&
  {Watson}}]{ued14}
{Ueda}, Y., {Akiyama}, M., {Hasinger}, G., {Miyaji}, T., \& {Watson},
  M.~G. 2014, \apj, 786, 104

\bibitem[{{Ueda} {et~al.}(2003){Ueda}, {Akiyama}, {Ohta}, \&
  {Miyaji}}]{ued03}
{Ueda}, Y., {Akiyama}, M., {Ohta}, K., \& {Miyaji}, T. 2003, \apj, 598, 886

\bibitem[{{Vignali} {et~al.}(2006){Vignali}, {Alexander}, \&
  {Comastri}}]{vign06qso2}
{Vignali}, C., {Alexander}, D.~M., \& {Comastri}, A. 2006, \mnras, 373, 321

\bibitem[{{Vignali} {et~al.}(2009){Vignali}, {Pozzi}, {Fritz}, {Comastri},
  {Gruppioni}, {Bellocchi}, {Fiore}, {Brusa}, {Maiolino}, {Mignoli}, {La
  Franca}, {Pozzetti}, {Zamorani}, \& {Merloni}}]{vign09qso2}
{Vignali}, C., et~al.\ 2009, \mnras,  395, 2189

\bibitem[{{Xue} {et~al.}(2011){Xue}, {Luo}, {Brandt}, {Bauer}, {Lehmer},
  {Broos}, {Schneider}, {Alexander}, {Brusa}, {Comastri}, {Fabian}, {Gilli},
  {Hasinger}, {Hornschemeier}, {Koekemoer}, {Liu}, {Mainieri}, {Paolillo},
  {Rafferty}, {Rosati}, {Shemmer}, {Silverman}, {Smail}, {Tozzi}, \&
  {Vignali}}]{xue11cdfs}
{Xue}, Y.~Q., et~al.\ 2011, \apjs, 195, 10

\bibitem[{{Zakamska} {et~al.}(2005){Zakamska}, {Schmidt}, {Smith}, {Strauss},
  {Krolik}, {Hall}, {Richards}, {Schneider}, {Brinkmann}, \&
  {Szokoly}}]{zaka05}
{Zakamska}, N.~L., et~al.\ 2005, \aj, 129, 1212

\bibitem[{{Zakamska} {et~al.}(2004){Zakamska}, {Strauss}, {Heckman},
  {Ivezi{\'c}}, \& {Krolik}}]{zaka04}
{Zakamska}, N.~L., {Strauss}, M.~A., {Heckman}, T.~M., {Ivezi{\'c}}, {\v Z}.,
  \& {Krolik}, J.~H. 2004, \aj, 128, 1002

\bibitem[{{Zakamska} {et~al.}(2003){Zakamska}, {Strauss}, {Krolik}, {Collinge},
  {Hall}, {Hao}, {Heckman}, {Ivezi{\'c}}, {Richards}, {Schlegel}, {Schneider},
  {Strateva}, {Vanden Berk}, {Anderson}, \& {Brinkmann}}]{zaka03}
{Zakamska}, N.~L., et~al.\ 2003, \aj, 126,  2125


\end{thebibliography}
\end{document}